\documentclass[12pt]{article}

\usepackage{lscape}
\usepackage{hyperref}
 \usepackage{enumitem}
\usepackage{bm} 

\usepackage{amsmath,amssymb,graphicx,multirow,textcomp,color}
\usepackage{epsfig,dcolumn,float,setspace}
\usepackage[text={7.1in,8in}, top=1.75in, left=0.69in]{geometry}
\usepackage{fancyvrb}

\begin{document}

\title{A regression-based method for detecting publication bias in multivariate meta-analysis}
\vspace{10mm}
\author{Chuan Hong\footnote{Department of Biomedical Informatics, Harvard Medical School, MA, USA}, Jing Zhang\footnote{Department of Epidemiology and Biostatistics, School of Public Health at University of Maryland, MD, USA}, Yang Li\footnote{Center for Statistical Science, Tsinghua University,  China}, Eleni Elia\footnote{Department of Biostatistics, Harvard T. H. Chan School of Public Health, MA, USA}, Richard Riley\footnote{Research Institute for Primary Care \& Health Sciences, Keele University, Staffordshire, UK} and Yong Chen\footnote{Department of Biostatistics, Epidemiology \& Informatics, University of Pennsylvania, PA, USA}}

\date{}
\maketitle
\vspace{10mm}
Publication bias occurs when the publication of research results depends not only on the quality of the research but also on its nature and direction. The consequence is that published studies may not be truly representative of all valid studies undertaken, and this bias may threaten the validity of systematic reviews and meta-analyses - on which evidence-based medicine increasingly relies. Multivariate meta-analysis has recently received increasing attention for its ability reducing potential bias and improving statistical efficiency by borrowing information across outcomes. However, detecting and accounting for publication bias are more challenging in multivariate meta-analysis setting because some studies may be completely unpublished whereas some studies may selectively report part of multiple outcomes. In this paper, we propose a score test for jointly testing publication bias for multiple outcomes, which is novel to the multivariate setting. The proposed test  is a natural multivariate extension of the univariate Egger's test, and can handle the above mentioned scenarios simultaneously, It accounts for correlations among multivariate outcomes, while allowing different types of outcomes, and can borrow information across outcomes. The proposed test is shown to be more powerful than the Egger's test, Begg's test and Trim and Fill method through simulation studies. Two data analyses are given to illustrate the performance of the proposed test in practice.
\vspace{5mm}
\newpage
\setlength{\baselineskip}{23pt} \vspace{-2mm}

\section{Introduction}
The interest in research synthesis and meta-analysis has rapidly grown over the last few decades. From today's point of view, it is difficult to think of scientific research without the possibility to integrate different findings into one big picture. In meta-analysis, publication bias (PB) is a well-known important and challenging issue may threaten the the validity of the results \cite{song2000publication, sutton2000empirical}.  PB is defined as the publication or non-publication of studies depending on the direction, the statistical significance of the results and other potential information such as language (selective inclusion of studies published in English), availability (selective inclusion of studies easily accessible to the researcher) and others \cite{rothstein2006publication}. It is important to differentiate PB from the inclusion bias due to the conduct of meta-analysis. For the latter bias, the criteria for including studies in a meta-analysis is influenced by knowledge of the results of the set of potential studies \cite{egger1998bias}. In this paper, we focus only on detecting PB and do not consider the inclusion bias due to the conduct of meta-analysis.

It may lead to erroneous conclusions being drawn from a meta-analysis if PB is not detected and adjusted. For example, the conclusions from some meta-analyses were later found to be contradicted with mega trials \cite{egger1995misleading}. In the last two decades, a great deal of attention has been devoted to developing statistical methods to detect and correct for PB using the data to be meta-analyzed \cite{rothstein2006publication, burkner2014testing}. Graphic methods based on funnel plots have been used as the basis to detect PB in meta-analysis. Because the precision in estimating the underlying treatment effect will increase as the sample size of component studies increases, results from small studies will scatter widely at the bottom of the funnel graph, with the spread narrowing among larger studies. A symmetrical inverted funnel indicates the absence of PB, while asymmetry inverted funnel indicates the potential PB \cite{light1984summing, sterne2001funnel}. However, observations based on funnel plots itself can be subjective. Without quantitatively measured the funnel plot asymmetry, different observers may have different interpretation of the observations. Statistical tests based on funnel plot asymmetry have been developed in the literature to assess PB, e.g., the rank correlation test \cite{begg1994operating}, and regression tests \cite{egger1997bias, macaskill2001comparison, harbord2006modified, peters2006comparison, rucker2008arcsine}. \cite{duval2000nonparametric, duval2000trim} further developed the ``Trim and Fill'' method for estimating and adjusting for the number of missing studies in meta-analysis. It is based on the idea that missing studies due to PB implies an asymmetrical funnel. The Trim and Fill method does not only test for PB, but also offers a correction to the estimates of overall effect size.

Recently, multivariate meta-analysis (MMA) has received a great deal of attention, which jointly analyzes multiple and possibly correlated outcomes \cite{jacksonmultivariate}. However, to the best of our knowledge, the commonly used graphical and statistical methods proposed for univariate outcome are no longer sufficient. For example, consider a meta-analysis with two primary outcomes. All existing univariate tests can only detect the PB of each outcome separately. One possible way to apply these tests in MMA is to define an combined univariate effect measure (e.g., diagnostic odds ratio combining the information on sensitivity and specificity). \cite{burkner2014testing} reviewed the performance of several statistical tests for PB in meta-analysis of diagnostic accuracy studies by applying the methods to univariate measures such as logarithm of diagnostic odds ratio, and concluded that using the Trim and Fill method on the logarithm of diagnostic odds ratio has non-inflated or slightly inflated Type I errors and medium to high power. However, the multivariate nature is not fully accounted for and there could be loss of information due to the reduction to one single measure \cite{burkner2014testing}. More importantly, for general meta-analysis with multivariate outcomes, the outcomes may be in different scales, e.g., meta-analysis of drug efficacy and safety studies \cite{amori2007efficacy}, and it is difficult to define a meaningful combined univariate effect measure (see section 5.3 for an example). Even if the outcomes are in the same scale, the interpretation of such a combined measure may be difficult.

In this paper, we propose regression-based test to detect PB in multivariate meta-analysis, which is a natural multivariate extension of the Egger's regression test. By jointly modeling on multivariate outcomes, the proposed test can fully account for the multivariate nature. In addition, by borrowing information across outcomes, this score test detects PB for both completely missing and partially missing scenarios simultaneously. Furthermore, the proposed test allows different types of outcomes. The proposed test approximately follows a simple $\chi^2$ distribution, and is robust and powerful comparing to existing univariate methods, suggesting that the proposed test can be used as a default method for test of PB.

This paper is organized as follows. In Section 2, we introduce the notations for multivariate random-effects meta-analysis and briefly review the existing methods for univariate meta-analysis based on funnel plots. In Section 3, we propose the regression-based score test. In Section 4, we conduct simulation studies to compare the proposed test with existing methods. We illustrate the proposed method by two real data analyses in Section 5. Finally, we provide a brief discussion in Section 6.

\section{Notations and existing methods}\label{sec:model}
\subsection{Notations for bivariate random-effects meta-analysis}
To simplify our presentation, we describe the bivariate random-effect model (BRMA) under normal distribution assumption, acknowledging that the proposed test can be easily extended to the multivariate setting. We consider a meta-analysis with $m$ studies where two outcomes in each study are of interest. For the $i$th study, denote $Y_{ij}$ and $s_{ij}$ the summary measure for the $j$th outcome of interest and associated standard error respectively, both assumed known, $i=1, \ldots, m$, and $j=1,2$. Each summary measure $Y_{ij}$ is an estimate of the true effect size $\theta_{ij}$. To account for heterogeneity in effect size across studies, we assume $\theta_{ij}$ to be independently drawn from a common distribution with overall effect size $\beta_j$ and between study variance $\tau_j^2$, $j=1,2$. Under normal distribution assumption for $Y_{ij}$ and $\theta_{ij}$, the general BRMA model can be written as \cite{van2002advanced}\\{{
\begin{eqnarray}\label{eq:brma}
\left(\begin{array}{c}Y_{i1} \\ Y_{i2} \end{array} \right)&\sim& N\left( \left(\begin{array}{c}\theta_{i1} \\ \theta_{i2} \end{array} \right), {\pmb{\Delta_i}}\right),\quad
{\pmb{\Delta_i}}=\left(\begin{array}{cc} s_{i1}^2 & s_{i1}s_{i2}\rho_{\textrm{W}i}\\ s_{i1}s_{i2}\rho_{\textrm{W}i} & s_{i2}^2 \end{array} \right),\nonumber
\\
\left(\begin{array}{c}\theta_{i1} \\ \theta_{i2} \end{array} \right)&\sim& N\left(\left(\begin{array}{c}\beta_{1} \\ \beta_{2} \end{array} \right), {\pmb{\Omega}}\right),\quad
{\pmb{\Omega}}=\left(\begin{array}{cc} \tau_{1}^2 & \tau_{1}\tau_{2}\rho_{\textrm{B}}\\ \tau_{1}\tau_{2}\rho_{\textrm{B}} & \tau_{2}^2 \end{array} \right),
\end{eqnarray}
where ${\pmb{\Delta_i}}$ and ${\pmb{\Omega}}$ }}are the corresponding study-specific within-study and between-study covariance matrices, and $\rho_{\textrm{W}i}$ and $\rho_{\textrm{B}}$ are the respective within-study and between-study correlations. When the within-study correlations $\rho_{\textrm{W}i}$ are known, inference on the overall effect sizes
$\beta_1$ and $\beta_2$, or their comparative measures (e.g. $\beta_1-\beta_2$), can be based on the marginal distribution of $(Y_{i1}, Y_{i2})$
{{
\begin{equation*}
\left(\begin{array}{c}Y_{i1} \\ Y_{i2} \end{array} \right)\sim N\left(\left(\begin{array}{c}\beta_{1} \\ \beta_{2} \end{array} \right), \bf{V_i}\right),
\bf{V_i}={\pmb{\Delta_i}}+{\pmb{\Omega}}=\left(\begin{array}{cc} s_{i1}^2+\tau_{1}^2 & s_{i1}s_{i2}\rho_{wi}+\tau_{1}\tau_{2}\rho_{\textrm{B}}\\ s_{i1}s_{i2}\rho_{wi}+\tau_{1}\tau_{2}\rho_{\textrm{B}} & s_{i2}^2+\tau_{2}^2 \end{array} \right).
\end{equation*}}}
We note that the variance of $Y_{ij}$ is partitioned into two parts, $s_{ij}^2$ and $\tau_j^2$, and the covariance, ${\textrm{cov}}(Y_{i1}, Y_{i2})=s_{i1}s_{i2}\rho_{wi}+\tau_{1}\tau_{2}\rho_{\textrm{B}}$, is also partitioned into two parts as the sum of within and between study covariances.

\subsection{Existing methods for PB}
The commonly used graphical methods can only handle PB for univariate meta-analysis. One of the most important graphical methods is the funnel plot \cite{light1984summing}. The most precise estimates (typically, those from the largest studies) are at the top of the funnel and those from less precise or smaller studies are at the base of the funnel. The precisions of the studies are plotted against summary statistics, whose shape should be like a funnel in the absence of PB \cite{duval2000trim}. If the studies with statistically significant results are more likely to be published, the shape of the funnel plot may become skewed or asymmetric.

Based on the funnel plot asymmetry, Egger (1997) \cite{egger1997bias} used a linear regression approach to detect PB. Specifically, Egger (1997) \cite{egger1997bias}  suggested to regress the effect size divided by its standard error as dependent variable $Y$ on the precision of the effect size $P$. In case of no publication bias, small studies are expected to be close to zero on both axes (due to their high standard error), whereas larger studies have a high precision and are thus expected to deviate from zero. In this case, the regression line is assumed to go through the origin, so that a does not differ significantly from zero. However, if publication bias is existent, most small studies will have relatively large effect size, and the intercept $a$ is thus assumed to be significantly greater than zero. In addition, each study can be weighted by the inverse of the variance under a fixed or random effects assumption to control for possible heteroscedasticity.
There are several variations of the Egger's regression test. \cite{macaskill2001comparison} introduced a different regression approach to detect PB, where the effect size is regressed on sample size of the study with weights being the inverse of the variances. A similar approach was suggested by \cite{peters2006comparison}, where the sample size is replaced by its inverse.

Begg and Mazumdar (1994) \cite{begg1994operating} proposed a non-parametric rank correlation method for PB based on Kendall's $\tau$, which tests whether the standardized effect size and the variance of the effect size are significantly associated. In the absence of publication bias, the variance should be independent of the effect measure, and thus, the test statistic is assumed to be close to zero. In the presence of publication bias, a test statistic that is significantly greater than zero indicates the presence of publication bias. Possible variations of Begg's rank test can be obtained by replacing standard errors by the inverse of total sample size.

Besides Begg's rank test, trim and fill method is another non-parametric method to detect and correct publication bias. It is based on the idea that there are $k$ studies present in the meta-analysis and $k_0$ studies missing due to publication bias, which implies an asymmetrical funnel. With respect to trim and fill, funnel plots are applied with an effect size on the x-axis and its precision on the y-axis. If we assume to know the true overall effect size, we will be able to estimate $k_0$.  The authors of trim and fill proposed three estimator (i.e., $R$, $L$ and $G$) of the number of missing studies (for details, see \cite{duval2000nonparametric, duval2000trim}). Using an iterative algorithm, one arrives at a random effects estimator that is used in the preceding formula. In the absence of publication bias (i.e., $k_0=0$), the approximate distributions of three estimator are known, and can be tested for significance to decide whether publication bias is present or not.

\section{Regression-based score test}
In this section, we propose a regression-based score test for PB, which is a multivariate extension of the Egger's regression test.
We consider a score test rather than a Wald test or a likelihood ratio test because the score test has the advantage of being obtained by fitting the null model only and being invariant to reparameterization.  The score test can be constructed using a three-step procedure: estimate the within-study correlation and variation; construct the regression model and likelihood; and propose the score test. 
\begin{enumerate}
\item The between-study variation $\tau^2$ and between-study correlation $\rho_{\textrm B}$ are estimated by maximizing the restricted maximum likelihood (REML) of Model (\ref{eq:brma}), denoted by $\tilde \tau^2$ and $\tilde \rho_{\textrm B}$. 

\item The bivariate regression model is constructed as
\begin{eqnarray}\label{eq:fl}
\textrm{\bf SND}_{i}&=&{\bf a}+{\bf b}\textrm {\bf P}_{i}+{\bm \varepsilon}_{i}
\end{eqnarray}
where $\textrm{\bf SND}_{i}=\left\{Y_{i1}\left(s_{i1}^2+\tau_2^2\right)^{-1/2}, Y_{i2}\left(s_{i2}^2+ \tau_2^2\right)^{-1/2}\right\}^T, \textrm {\bf P}_{i}=\left\{\left(s_{i1}^2+ \tau_1^2\right)^{-1/2}, \left(s_{i2}^2+ \tau_2^2\right)^{-1/2}\right\}^T,$ and 
${\bm \varepsilon}_{i}=\left\{\varepsilon_{i1},\varepsilon_{i2}\right\}^T$ is standard bivariate normal random variables with covariance matrix ${\pmb \Sigma}_i$, and ${\pmb\Sigma}_i$ is a $2 \times 2$ matrix with 1 as the diagonal elements and $\left(\rho_{\textrm{W}_i}s_{i1}s_{i2}+\rho_{\textrm B}\tau_1\tau_2\right)/ \sqrt{\left(s_{i1}^2+\tau_1^2\right)\left(s_{i2}^2+\tau_2^2\right)}$ as the off-diagonal elements. 
\\
\\
By plugging in  $\tilde \tau^2$ and $\tilde \rho_{\textrm B}$ obtained from step (I), the loglikelihood is then constructed as 
\begin{equation}\label{eq:lik}
\log L(a, b)=-{1\over 2}\sum_{i=1}^m {(\widetilde{\textrm{SND}}_{i}-a-b \tilde{\textrm{P}}_{i})^2} ,
\end{equation}
\item The score test based on model (\ref{eq:lik}) is proposed, referred to as RST hereafter,
\begin{equation}
\label{RST}{\textrm{RST}}={1\over m}  {\bf U}_{a}\left[{\bf 0}, \tilde {\bf b}({\bf 0})\right]^T{{\bf I}_0}^{aa}{\bf U}_{a}\left[{\bf 0}, \tilde {\bf b}({\bf 0})\right],
\end{equation}
where ${\bf U}_{a}\left[{\bf 0}, \tilde {\bf b}({\bf 0})\right]$ is the score function w.r.t $\bf a$ under the null, and ${{\bf I}_0}^{aa}$ is sub-matrix of inverse of the negative hessian function evaluated at $(0,\tilde {\bf b}({\bf 0}))$ w.r.t $\bf a$.
\end{enumerate}
A detailed derivation of the score function and test statistics is provided in Appendix A of supplementary materials.  {\color{red}To be added: I am still working on the appendix}

\section{Simulation studies}
{In this section, we evaluate the performance of the proposed RST test through fully controlled simulation studies.}
\subsection{Simulation studies}
{In our simulation studies, we compare the proposed RST test with the commonly used Egger's regression test, the Begg's rank test and the Trim and Fill method.} The data are generated from BRMA as specified by model (\ref{eq:brma}). To cover a wide spectrum of scenarios, we vary the values for several factors that are considered important in practice: 1)To reflect the heterogeneity in standard error of summary measure across studies, we sample $s_{ij}^2$ from the square of $N(0.3, 0.50)$ distribution, which leads to a mean value of $0.33$ for $s_{ij}^2$; 2) The size of the within-study variation relative to the between-study variation may have a substantial impact on the performance of the methods. To this end, we let the between-study variances $\tau_1^2=\tau_2^2$ ranging from 0.5 to 1.9 to represent the random-effects model with relatively small to large random effects. 3) For within-study and between-study correlations, we consider $\rho_{\textrm{W}i}$ being constant with value $-0.5$, $0$ or $0.5$ and $\rho_{\textrm{B}}$ being constant with $-0.5$, $0$ or $0.5$. 4) The number of studies $n$ is set to 50, 75 and 100 to represent meta-analysis of small to large number of studies.

We consider the Type I error setting when the {selection (i.e., a study being published or not)} does not depend on the effect sizes, and power settings when the selection depends on the effect sizes.  We conduct 5000 simulations for the Type I error setting, and 1000 simulations for the power settings. For power settings, in order to obtain $n$ published studies, we follow the three steps: 1) $N$ studies are simulated, where $N$ is an positive integer greater than $n$ (here we choose $N=3n$ to ensure enough number of studies before selection); 2) studies are excluded based on the selection scenario; 3) finally, $n$ studies were randomly sampled from the studies included in the previous step. We consider missing scenarios for completely missing and partially missing. The probability of an outcome being reported depends on the p value of its effect size, therefore some studies may selectively report only one of the outcomes. In addition, the selection model in the partially missing scenario is different from those in completely missing scenarios, in that we let the logit of the probability of the $j$th outcome in $i$th study being published be $(-2.5+0.1\textrm{SND}_{ij}+1.5\textrm{SND}_{ij}^2)I(\textrm{SND}_{ij}<2)+4I(\textrm{SND}_{ij}\ge2)$. This selection model is empirically estimated from a meta-analysis where the true status of a registered study being published or not is available \cite{turner2008selective}.

\subsubsection{Simulation Results}
Table~\ref{tab:sim} summarizes the Type I errors at the $10\%$ nominal level of the tests under comparison. The proposed RST test controls the Type I errors well in all settings. All univariate tests are based on combined univariate effect measures of two outcomes. The Egger's regression test has slightly inflated Type I errors when the sample size is small ($n=10$). We observe that the Type I errors of the Begg's rank test are very conservative when the sample size is relatively small but inflated when the sample size is relatively large, which is consistent with literature in that the Begg's rank test does not perform well in controlling Type I errors \cite{burkner2014testing}. We observe that the Trim and Fill method controls the Type I errors well only when the between-study heterogeneity is relatively small and is too conservative when the between-study heterogeneity is relatively large. The possible reason is that the Trim and Fill method is very sensitive to outliers, and larger between-study heterogeneity may introduce more outliers.

\begin{figure}[!h]
\centering\includegraphics[scale=0.8]{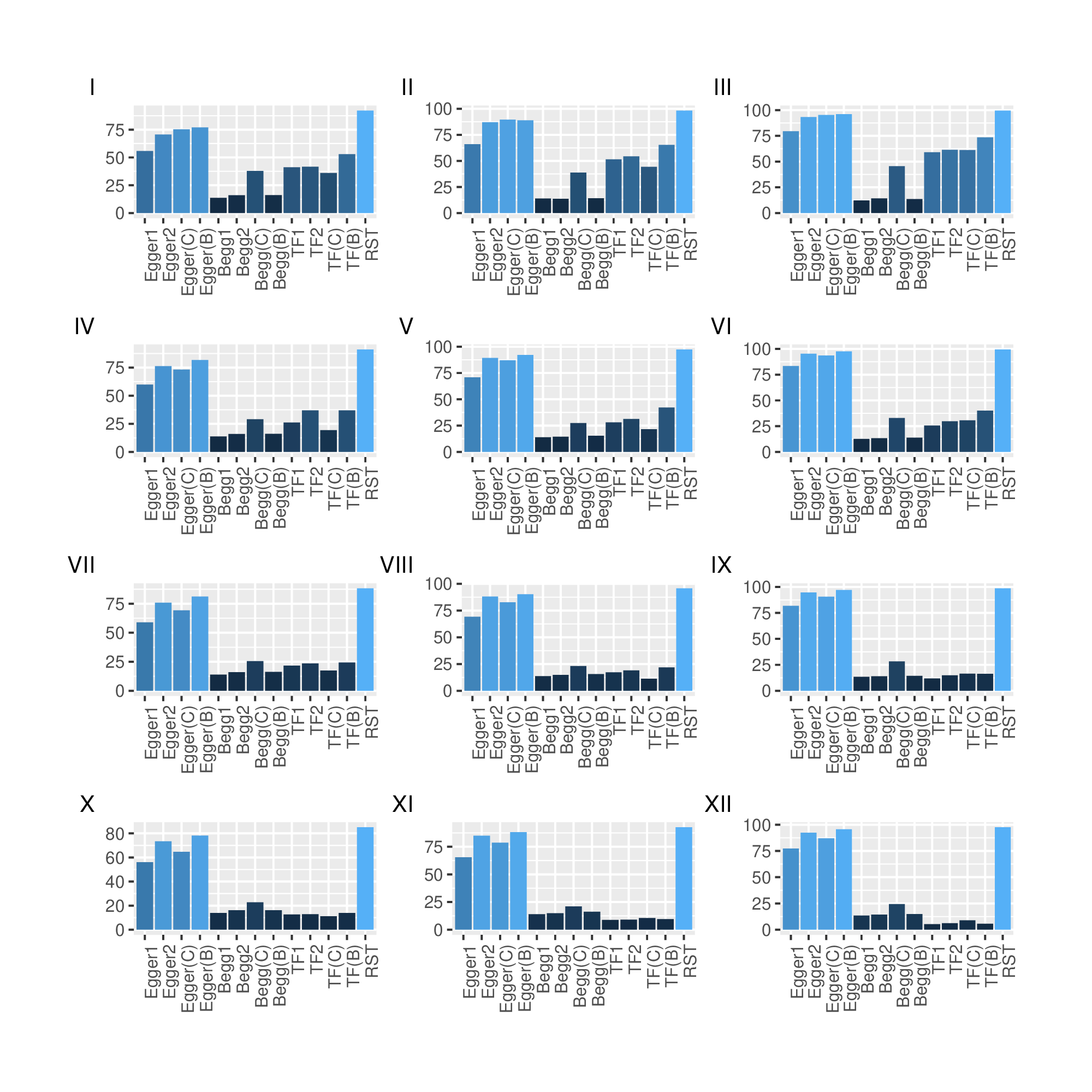}
\caption{Power plot. (will change the panel title later)}
\label{fig:power}
\end{figure}

Figure~\ref{fig:power} summarizes the power of the tests under comparison for complete missing scenario. We display the power curve by adjusting for its critical region. Clearly, the proposed RST test is the most powerful test under all settings considered. The Begg's rank test and the Trim and Fill method are essential no power beyond the nominal level when the between-study variance is large. There are several interesting findings from Figure~\ref{fig:power}.   We observe decreasing trend for the Begg's rank test and the Trim and Fill method when between-study variance is increasing. This is reasonable because both tests are nonparametric tests based on ranks, which are sensitive to outliers. Larger heterogeneity will introduce severer problem caused by outliers.

\begin{table}[!h]
\centering
\caption{Type I errors ($\times 100\%$) at $10\%$ nominal level of Egger test, Begg test, Trim and Fill method and the proposed RST test. The univariate tests based on the combined univariate measure or Bonferroni correction are denoted as ``test(C)'' or ``test(B)''.  The number of studies $n$ is set to 50, 75 and 100, and the between-study heterogeneity $\tau^2$ is set to 0.5, 1.1, 1.5 and 1.9}
{\small
\begin{tabular}{ccccccccccccccc}
\hline
$\tau^2$ & n   & Egger$_1$ & Egger$_1$ & Egger(C) & Egger(B) & Begg$_1$ & Begg$_1$ & Begg(C) & Begg(C) & TF$_1$  & TF$_1$ & TF(C) & TF(B)& RST  \\
\hline
0.5   & 50  & 10.3   & 10.2   & 10.7           & 10.5              & 43    & 43.2  & 13.2          & 59.2             & 11.6 & 10.8 & 11.1        & 12.7           & 13.3 \\
      & 75  & 10.4   & 10.1   & 10.2           & 10.2              & 53.1  & 54    & 17.4          & 71.2             & 11.5 & 11.2 & 10.9        & 12.6           & 12.6 \\
      & 100 & 10.2   & 10.8   & 9.8            & 10                & 59.5  & 58.1  & 21.7          & 77.4             & 10.7 & 11.3 & 10.2        & 12             & 12.2 \\
      \\
1.1   & 50  & 10.3   & 10     & 10.6           & 10.3              & 44.6  & 44.6  & 14            & 61.1             & 10.5 & 9.6  & 9.8         & 11.8           & 12.7 \\
      & 75  & 10.6   & 10.2   & 10             & 10.2              & 54.2  & 55.4  & 18.7          & 72.9             & 10   & 10   & 9.7         & 11.5           & 12.1 \\
      & 100 & 10.2   & 10.7   & 9.9            & 10.1              & 60.8  & 59.4  & 23            & 78.9             & 9.9  & 10.3 & 9.7         & 11.4           & 11.4 \\
      \\
1.5   & 50  & 10.2   & 10.1   & 10.5           & 10.3              & 44.9  & 45.1  & 14.2          & 61.6             & 9.8  & 9.2  & 9.3         & 11.3           & 12.6 \\
      & 75  & 10.5   & 10.3   & 9.9            & 10.2              & 54.6  & 55.7  & 19.1          & 73.4             & 9.4  & 9.6  & 9.3         & 10.9           & 11.9 \\
      & 100 & 10.3   & 10.7   & 9.9            & 10                & 61.1  & 59.6  & 23.3          & 79.2             & 9.5  & 9.6  & 9.5         & 11             & 11.4 \\
      \\
1.9   & 50  & 10.2   & 10.1   & 10.6           & 10.2              & 45.2  & 45.3  & 14.3          & 61.9             & 9.4  & 8.8  & 8.9         & 10.9           & 12.4 \\
      & 75  & 10.5   & 10.3   & 9.9            & 10.3              & 54.8  & 56    & 19.2          & 73.6             & 9.1  & 8.9  & 8.9         & 10.4           & 11.8 \\
      & 100 & 10.3   & 10.7   & 10             & 10.1              & 61.3  & 59.8  & 23.5          & 79.4             & 9.1  & 9.2  & 9.2         & 10.7           & 11.3\\
\hline
\end{tabular}
}
\label{tab:sim}
\end{table}

\section{Data analysis}
In this section, we consider two case studies: 1) lymph node metastasis in women with cervical cancer; 2) efficacy and safety of drug-eluting stens and bare-metal stens.  The first analysis illustrates the situation when the existing univariate tests can be applied to a combined outcome, but the combination procedure of two outcomes will hamper the testing power. The second analysis illustrate the situation when two outcome cannot be combined hence the existing univariate tests can be only applied separately to the outcomes. Under both situations, the proposed score test shows significant power gain over the existing univariate tests. 
\subsection{Diagnosis of lymph node metastasis in women with cervical cancer}
We compare the performance of the proposed test with existing univariate tests by detecting the publication bias of the data in \cite{scheidler1997radiological}. Scheidler et al. compared three imaging techniques for the diagnosis of lymph node metastasis in women with
cervical cancer. Forty-four studies in total were included: 17 studies evaluated lymphangiography, another 17 studies
examined computed tomography and the remaining 10 studies focused on magnetic resonance imaging. Diagnosis of
metastatic disease by lymphangiography (LAG) is based on the presence of nodal-filling defects, whereas computed tomography (CT) and magnetic resonance imaging (MRI) rely on nodal enlargement.

The results of the statistical tests are shown in the upper panel of Table~\ref{tab:case studies}. The proposed RST test provides significant results by jointly estimating publication bias for sensitivities and specificities. Both the Egger test and Begg rank test detect significant publication bias for specificity. While severe asymmetry has been observed for specificities, the Trim and Fill method does not detect publication bias for neither sensitivity nor specificity. A possible reason for this result is that the Trim and Fill method is very sensitive to outliers. Using the diagnostic odds ratio as a combined measure to detect publication bias may lead to loss of power, and non of the three univariate methods identify significant publication bias based on logDOR.

\begin{table}[!h]
\centering
\caption{P values of testing publication bias using the Egger test, the Begg test, the Trim and Fill method (TF) and the proposed omnibus score test (RST).}
\begin{tabular}{ccccc}
\hline
           &       Test & Outcome 1 & Outcome 2 &     Combined univariate measure \\
\hline
Case study 1 &      Egger &      0.642 &      $<0.001$ &      0.091 \\

           &       Begg &      0.761 &      0.002 &      0.511 \\

           &         TF &      0.125 &      0.250 &      0.500 \\
           &       MPBT &      \multicolumn{2}{c}{$<0.001$ }          &            \\
           \\
Case study 2 &      Egger &      0.933 &      0.619 &     --\\

           &       Begg &      1.00 &      0.720 &      -- \\

           &         TF &      0.500 &      0.500 &     -- \\
           &       MPBT &      \multicolumn{2}{c}{0.006}          &            \\
\hline
\multicolumn{5}{l}{$^*$ jointly test for publication bias of Outcome 1 and Outcome 2}\\
\end{tabular}
\label{tab:case studies}
\end{table}

\subsection{Efficacy and safety of drug-eluting stents vs. bare-metal stents in acute myocardial infarction}
Primary percutaneous coronary intervention (PCI) is a preferred reperfusion strategy for patients presenting with
acute myocardial infarction. Routine implantation
of bare-metal stents has been associated with improved clinical outcome mainly because of the decreased risk for
reintervention. Restenosis is an important limitation of the use of bare-metal stents in
patients with acute myocardial infarction. Drug-eluting stents effectively reduce restenosis while
maintaining a good safety profile in many lesion and patients groups. However, concerns have been raised with regard to the safety of drug-eluting stents in patients with acute myocardial infarction. Implantation of drug-eluting stents during
primary PCI could be associated with an increased risk for stent thrombosis. Recent studies reported results of drug-eluting stents in patients undergoing primary PCI for acute myocardial infarction. However, the power of these studies were insufficient to assess the rare adverse events, and the results were  not consistent. \cite{kastrati2007meta} performed a meta-analysis based on
individual patient data from randomized trials comparing drug-eluting stents with bare-metal stents to evaluate the efficacy and safety of drug-eluting stents. Eight randomized trials comparing drug-eluting
stents with bare-metal stents in 2786 patients with acute
myocardial infarction were included in the meta-analysis. \cite{kastrati2007meta}  concluded that the use of drug-eluting stents in patients with acute myocardial infarction was safe and improved clinical outcomes by reducing the risk of reintervention compared
with bare-metal stents.

We estimate the publication bias of the primary efficacy outcome and primary safety outcome using the Egger test, the Begg test, the Trim and Fill method and the proposed MPBT test. It is worth mentioning that unlike the meta-analysis of diagnostic accuracy study where a combined univariate measure (i.e., $\log$DOR) can be used for the existing univariate tests, for the meta-analysis of drug efficacy and safety study, it is hard to interpret its clinical meaning when combine the efficacy outcome and primary safety outcome together. Therefore, we apply the existing univariate tests on the efficacy outcome and safety outcome separately. On the other hand, the proposed MPBT has no such limitation and can test for publication bias for the two outcomes jointly.

The results of the statistical tests are shown in the lower panel of Table\ref{tab:case studies}. The proposed MPBT test provides a significant result by jointly estimating publication bias for the efficacy and the safety outcomes. Both the Egger test and Begg rank test detect significant publication bias for specificity. While severe asymmetry has been observed for specificities, the Trim and Fill method does not detect publication bias for neither sensitivity nor specificity.

\section{Discussion}
We have proposed an omnibus score test for detecting publication bias in multivariate random-effects meta-analysis for both completely unpublished scenario and selectively report part of multiple outcome scenario, which is a natural multivariate extension of the multivariate regression model. This approach has a variety of advantages.
	
In the proposed test the joint modeling on multiple outcomes uses more information and limits some of the challenges present in standard tests for publication bias. For example for insufficient number of studies (less than 10) there is not enough power for standard tests to distinguish chance from real asymmetry. The multivariate nature of the proposed test on the other hand, allows us to use correlated outcomes and thereby borrow information across outcomes and increase power. The proposed test also allows for jointly testing for multiple outcomes with different types and scales (as illustrated in the case studies) by borrowing information across outcomes.   

Another problem that this test addresses is the selective reporting/publishing of only significant results by investigators who may measure different types of outcomes or just a single outcome. These cases correspond to partially and completely missing scenarios respectively. The proposed test detects PB for both scenarios simultaneously. Specifically, in the simulation study the proposed test yielded results with better statistical properties, controlled for type I errors well and was shown to be more powerful than existing tests.   

One limitation of the proposed test is it requires within-study correlations to be known, which are often not available. Nonetheless, within-study correlations can be obtained, from potentially available individual patient data (IPD) by a proportion of studies, using bootstrapping methods. In the case where the IPD are not available, extensions of the proposed test can be constructed using alternative synthesis models that do not require within-study correlations(such as Riley's method). Bayesian extension of the proposed model can also be considered placing a prior distribution upon unknown within-study correlations. The extension of the proposed test using estimated within-study correlations is a topic of future work. 

To summarize, we have developed a useful test to detect publication bias in multivariate random-effects meta-analysis. By well designed simulation studies, we found the proposed test is substantially more powerful than the existing tests by borrowing information across outcomes. Thus, this test can be a useful addition to tackle the publication bias problem in comparative effectiveness research.

\bibliographystyle{authordate2}
\bibliography{mPB_flst}

\begin{thebibliography}{10}
\providecommand{\url}[1]{\texttt{#1}}
\providecommand{\urlprefix}{URL }
\expandafter\ifx\csname urlstyle\endcsname\relax
  \providecommand{\doi}[1]{doi:\discretionary{}{}{}#1}\else
  \providecommand{\doi}{doi:\discretionary{}{}{}\begingroup
  \urlstyle{rm}\Url}\fi

\bibitem{song2000publication}
Song F, Eastwood AJ, Gilbody S, Duley L, Sutton AJ. Publication and related
  biases. \emph{Health technology assessment (Winchester, England)}  2000;
  \textbf{4}(10):1.

\bibitem{sutton2000empirical}
Sutton AJ, Duval S, Tweedie R, Abrams KR, Jones DR, \emph{et~al.}. Empirical
  assessment of effect of publication bias on meta-analyses. \emph{Bmj}  2000;
  \textbf{320}(7249):1574--1577.

\bibitem{rothstein2006publication}
Rothstein HR, Sutton AJ, Borenstein M. \emph{Publication bias in meta-analysis:
  Prevention, assessment and adjustments}. John Wiley \& Sons, 2006.

\bibitem{egger1998bias}
Egger M, Smith GD. Bias in location and selection of studies. \emph{BMJ:
  British Medical Journal}  1998; \textbf{316}(7124):61.

\bibitem{egger1995misleading}
Egger M, Smith GD. Misleading meta-analysis. \emph{Bmj}  1995;
  \textbf{310}(6982):752--754.

\bibitem{burkner2014testing}
B{\"u}rkner PC, Doebler P. Testing for publication bias in diagnostic
  meta-analysis: a simulation study. \emph{Statistics in medicine}  2014;
  \textbf{33}(18):3061--3077.

\bibitem{light1984summing}
Light R, Pillemer DB. \emph{Summing up}. Harvard University Press, 1984.

\bibitem{sterne2001funnel}
Sterne JA, Egger M. Funnel plots for detecting bias in meta-analysis:
  guidelines on choice of axis. \emph{Journal of clinical epidemiology}  2001;
  \textbf{54}(10):1046--1055.

\bibitem{begg1994operating}
Begg CB, Mazumdar M. Operating characteristics of a rank correlation test for
  publication bias. \emph{Biometrics}  1994; :1088--1101.

\bibitem{egger1997bias}
Egger M, Smith GD, Schneider M, Minder C. Bias in meta-analysis detected by a
  simple, graphical test. \emph{Bmj}  1997; \textbf{315}(7109):629--634.

\bibitem{macaskill2001comparison}
Macaskill P, Walter SD, Irwig L. A comparison of methods to detect publication
  bias in meta-analysis. \emph{Statistics in medicine}  2001;
  \textbf{20}(4):641--654.

\bibitem{harbord2006modified}
Harbord RM, Egger M, Sterne JA. A modified test for small-study effects in
  meta-analyses of controlled trials with binary endpoints. \emph{Statistics in
  medicine}  2006; \textbf{25}(20):3443--3457.

\bibitem{peters2006comparison}
Peters JL, Sutton AJ, Jones DR, Abrams KR, Rushton L. Comparison of two methods
  to detect publication bias in meta-analysis. \emph{Jama}  2006;
  \textbf{295}(6):676--680.

\bibitem{rucker2008arcsine}
R{\"u}cker G, Schwarzer G, Carpenter J. Arcsine test for publication bias in
  meta-analyses with binary outcomes. \emph{Statistics in medicine}  2008;
  \textbf{27}(5):746--763.

\bibitem{duval2000nonparametric}
Duval S, Tweedie R. A nonparametric "trim and fill" method of accounting for
  publication bias in meta-analysis. \emph{Journal of the American Statistical
  Association}  2000; \textbf{95}(449):89--98.

\bibitem{duval2000trim}
Duval S, Tweedie R. Trim and fill: a simple funnel-plot--based method of
  testing and adjusting for publication bias in meta-analysis.
  \emph{Biometrics}  2000; \textbf{56}(2):455--463.

\bibitem{jacksonmultivariate}
Jackson D, Riley R, White I. Multivariate meta-analysis: Potential and promise.
  \emph{Statistics in Medicine}  2011; \textbf{30}(20):2481--2498.

\bibitem{amori2007efficacy}
Amori RE, Lau J, Pittas AG. Efficacy and safety of incretin therapy in type 2
  diabetes: systematic review and meta-analysis. \emph{Jama}  2007;
  \textbf{298}(2):194--206.

\bibitem{van2002advanced}
Van~Houwelingen HC, Arends LR, Stijnen T. Advanced methods in meta-analysis:
  multivariate approach and meta-regression. \emph{Statistics in medicine}
  2002; \textbf{21}(4):589--624.

\bibitem{turner2008selective}
Turner EH, Matthews AM, Linardatos E, Tell RA, Rosenthal R. Selective
  publication of antidepressant trials and its influence on apparent efficacy.
  \emph{New England Journal of Medicine}  2008; \textbf{358}(3):252--260.

\bibitem{scheidler1997radiological}
Scheidler J, Hricak H, Kyle KY, Subak L, Segal MR. Radiological evaluation of
  lymph node metastases in patients with cervical cancer: a meta-analysis.
  \emph{Jama}  1997; \textbf{278}(13):1096--1101.

\bibitem{kastrati2007meta}
Kastrati A, Dibra A, Spaulding C, Laarman GJ, Menichelli M, Valgimigli M,
  Di~Lorenzo E, Kaiser C, Tierala I, Mehilli J, \emph{et~al.}. Meta-analysis of
  randomized trials on drug-eluting stents vs. bare-metal stents in patients
  with acute myocardial infarction. \emph{European heart journal}  2007;
  \textbf{28}(22):2706--2713.

\end{thebibliography}
\end{document}